\documentclass[aps,twocolumn,showpacs]{revtex4}
\usepackage{graphicx}

\begin{document}
\draft

\newcommand{\beq}{\begin{equation}}
\newcommand{\eeq}{\end{equation}}
\newcommand{\ben}{\begin{eqnarray}}
\newcommand{\een}{\end{eqnarray}}
\newcommand{\bea}{\begin{array}}
\newcommand{\eea}{\end{array}}
\newcommand{\om}{(\omega )}
\newcommand{\bef}{\begin{figure}}
\newcommand{\eef}{\end{figure}}
\newcommand{\leg}[1]{\caption{\protect\rm{\protect\footnotesize{#1}}}}

\newcommand{\ew}[1]{\langle{#1}\rangle}
\newcommand{\be}[1]{\mid\!{#1}\!\mid}
\newcommand{\no}{\nonumber}
\newcommand{\etal}{{\em et~al }}
\newcommand{\geff}{g_{\mbox{\it{\scriptsize{eff}}}}}
\newcommand{\da}[1]{{#1}^\dagger}
\newcommand{\cf}{{\it cf.\/}\ }
\newcommand{\ie}{{\it i.e.\/}\ }

\title{\center{Room temperature stable single-photon source}}

\author{Alexios Beveratos, Sergei K\"uhn, Rosa Brouri, Thierry Gacoin$^*$,
Jean-Philippe Poizat and Philippe Grangier}
\affiliation{Laboratoire Charles Fabry de l'Institut  d'Optique, UMR 8501 du
CNRS, \\
B.P. 147, F-91403 Orsay Cedex - France. \\
$^*$ Laboratoire de Physique de la mati\`ere condens\'ee,
Ecole polytechnique, F-91128 Palaiseau, France.\\
{\rm e-mail : alexios.beveratos@iota.u-psud.fr}}

%
%
\vspace{1cm}

\begin{abstract}

We report on the realization of a stable solid state room temperature 
 source for single photons. It is based on the fluorescence of a single
nitrogen-vacancy (NV) color center in
a diamond nanocrystal. Antibunching has been observed 
in the  fluorescence light under both continuous and pulsed excitation.
Our source delivers $2\times 10^4$ s$^{-1}$ single-photon pulses at an excitation repetition
rate of $10$ MHz. The number of two-photon pulses is reduced by a factor of five compared to
strongly attenuated coherent sources. 

\end{abstract}


\maketitle

\section{Introduction}

The security of quantum cryptography 
is based on the
fact that quantum mechanics does not allow
one to
duplicate an unknown state of a single quantum system (for a review see \cite{TRG}).
This property is referred to as the no-cloning theorem.
After the pioneering experiment of the group of Bennett and Brassard
\cite{BBBSS}, several quantum key distribution
set-ups using attenuated laser pulses have been demonstrated 
(see for example refs \cite{TRG,BHLMNP}).
In these implementations, single photons are approximated by
strongly attenuated
coherent pulses so that the average number of photons per pulse is
$p_1\approx 0.1$. In this case, the probability
of having two photons in a pulse is approximately $p_1^2/2$ \cite{LoiPoisson}. Two-photon
pulses are a potential information leakage source \cite{L}.  Indeed, an eavesdropper could
tap on the communication between the sender (Alice) and the receiver (Bob), by keeping one
of the  two photons  and sending the other one to Bob. The eavesdropper can then
measure the state of the photon once Alice and Bob have revealed their
measurement basis. 
With attenuated coherent pulses, the only way to reduce the probability of
having two photons in a pulse is to lower $p_1$ and thereby decreasing 
the transmission rate. Our aim is to realize an efficient single-photon source
that would have a vanishing two photon probability for a non vanishing
transmission rate.

Single photons on demand can be produced by pulsed excitation of a single dipole
\cite{MGM,pra}. The
principle is that a single emitting dipole has to undergo a full
excitation-emission-reexcitation cycle before
emitting a second photon. For a sufficiently short and intense excitation pulse,
a single dipole emits one and only one photon
\cite{pra}.

Several solid state sources, like single organic
molecules \cite{BMOT,BLTO,KJRT,wild,ML,TCGR}, self-assembled semiconductor quantum
dots \cite{YY,I2,ZBJPJTGPSB}, or
semiconductor nanocrystals \cite{I1} have been presented lately as potential
candidates. However, the ideal candidate should be
photostable, work at room
temperature and be easy to manipulate. Single nitrogen-vacancy (NV)
 color centers in bulk diamond \cite{GDTFWB} have recently been 
 shown to  exhibit strong antibunching at room
temperature \cite{ol,capri,KMZW}. They are intrinsically photostable and 
are believed to have a unity quantum efficiency \cite{efficiency}. 
The
high refractive index of diamond leads however to small a collection efficiency owing
to total internal
reflection and spherical aberrations. Also, the signal to background ratio
is limited by the light
emitted from the surrounding diamond crystal. One can also point out that bulk
diamond cannot be manufactured in
any desirable shape, and thus is very difficult to handle and insert, for example,
into a microcavity.

In this paper, we show that the use of single NV color center in diamond
nanocrystals (typical size $50$ nm)
is a very convenient solution to these problems. The subwavelength size of
the nanocrystals renders refraction
irrelevant. One can simply think of the nanocrystal as a point source
emitting light in air.
Furthermore, the small volume of diamond excited by the pump light reduces
the emitted background
light. Also, diamond nanocrystals can be easily handled in order to be inserted in
a cavity or to be deposited on a fiber tip
\cite{KHSPV}.  In addition NV centers in diamond nanocrystals preserve all
the important features of
NV centers in bulk diamond. 
In particular, we have checked that
the fluorescence spectrum of NV centers in nanocrystals at room temperature is the same
as  in  bulk.
By investigating the autocorrelation function
under continuous wave (CW) and pulsed
excitation, we demonstrate the possibility to use  NV centers in diamond
nanocrystals as a room
temperature stable single-photon source.

\section{Experimental set-up}

The color center used in our experiments is the NV defect center in
synthetic Ib diamond, with a zero
phonon line at a wavelength of $637$ nm \cite{GDTFWB}. The defect consists
in a substitutional
nitrogen and a vacancy in an adjacent site. A simplified level structure is
a four-level scheme
with fast non radiative decays within the two upper states and within the
two lower states. This amounts to an incoherent two-level system.
The lifetime of the excited state in the bulk is $\tau_b =11.6$ ns \cite{CTJ}.

NV centers are  artificially created in synthetic MDA diamond powder from de Beers.
 Nitrogen is naturally
present in diamond. Vacancies are created by  irradiation with $1.5$ MeV
electrons at a dose of
$3\times 10^{17} e^{-}/$cm$^2$. Subsequent annealing in vacuum at
$850^o$C during $2$ hours leads to the formation of NV centers \cite{GDTFWB}. The
nanocrystals are  dispersed by sonification in a solution of polymer
(Polyvinylpyrrolidone at 1 weight$\%$ in propanol). This allows the
disaggregation of the particles
and their stabilization in a colloidal state. Centrifugation  at 11000 rpm
for 30 min allows us to
select nanocrystal sizes of $d_0=90\pm 30$ nm (measured by dynamical light
scattering). The average
number of NV centers in a nanocrystal has been evaluated to $8$ (see
below). Nanocrystals containing
a single NV center should  therefore have a size around $d_0 /2 =45$ nm,
 which lies  in the lower
wing of the size distribution.  The nanocrystal solution is then  spin
coated at 3000 rpm on thin
fused silica substrates. Evaporation of the solvent leaves a 30 nm thick
film of polymer with the
nanocrystals well dispersed on the surface. Their density was estimated to
be around  $0.02$ $\mu$m$
^{-2}$. In most experiments we look at the centers from the other side of the
plate, which is in
contact with the oil of an immersion microscope lens (Nachet 004279, N.A. =
1.3).

\begin{figure} 
\includegraphics[scale=0.32]{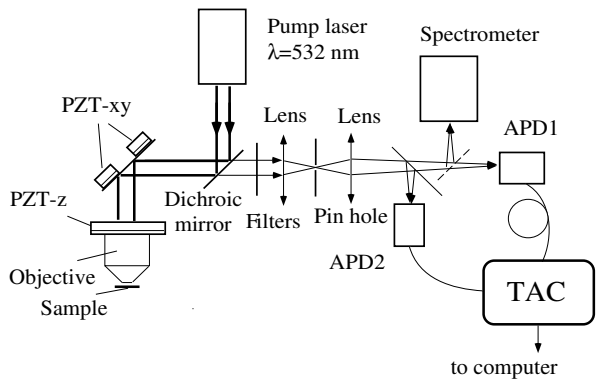}
\caption{Experimental set-up. The sample is excited by either a continuous or a pulsed
frequency doubled YAG. Its fluorescence is collected using a confocal microscope
set-up. The intensity correlations are measured using two avalanche photodiodes on each side
of a 50/50 beam splitter, a time to amplitude converter and  a multichannel analyzer.}
\label{expsetup}
\end{figure}

The experimental setup (fig. \ref{expsetup}), based on a confocal microscope, has
been described in detail elsewhere \cite{ol,capri}. The pump laser (CW or
pulsed) is focused
with an immersion oil, high numerical aperture objective ($NA=1.3$), onto a diffraction
limited spot ($\approx 400$ nm FWHM). The
fluorescence light is collected by the same objective and after proper
frequency and spatial filtering, it can either
be send on a spectrometer, or to a Hanbury-Brown and Twiss set-up using
two avalanche photodiodes (APD) from EG$\&$G. Appropriate data processing allows us to
obtain the histogram of the time separations between successive photons.
A slow (8s response time) x-y-z computerized servo-loop is used to compensate for any drifts.
Bandpass filters allow detection from $630$ to $800$ nm. 
This spectral window matches  the broad emission spectrum 
of a NV center \cite{GDTFWB}.

\section{Continuous wave excitation}

\begin{figure} 
\includegraphics[scale=0.3]{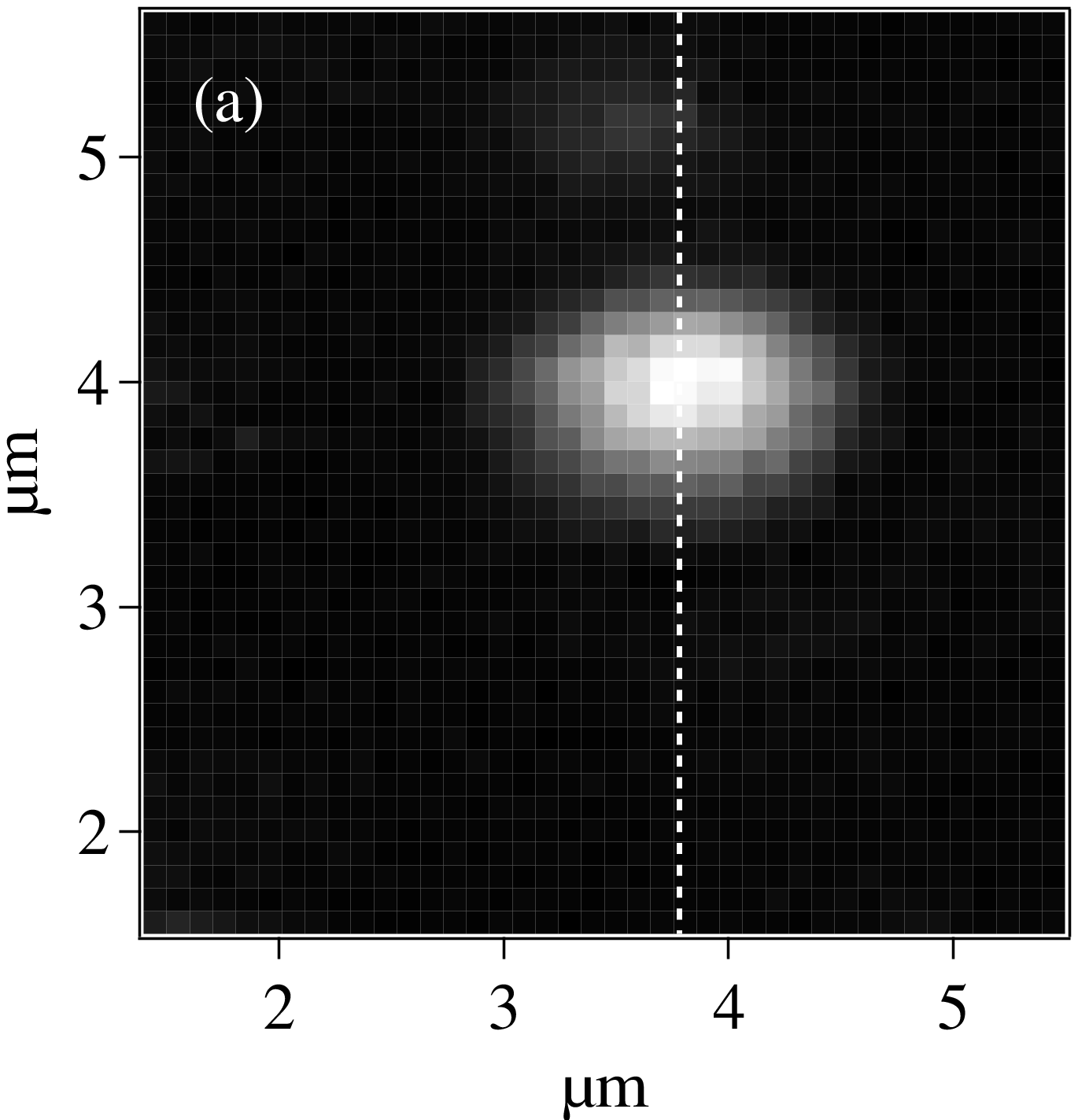}
\includegraphics[scale=0.3]{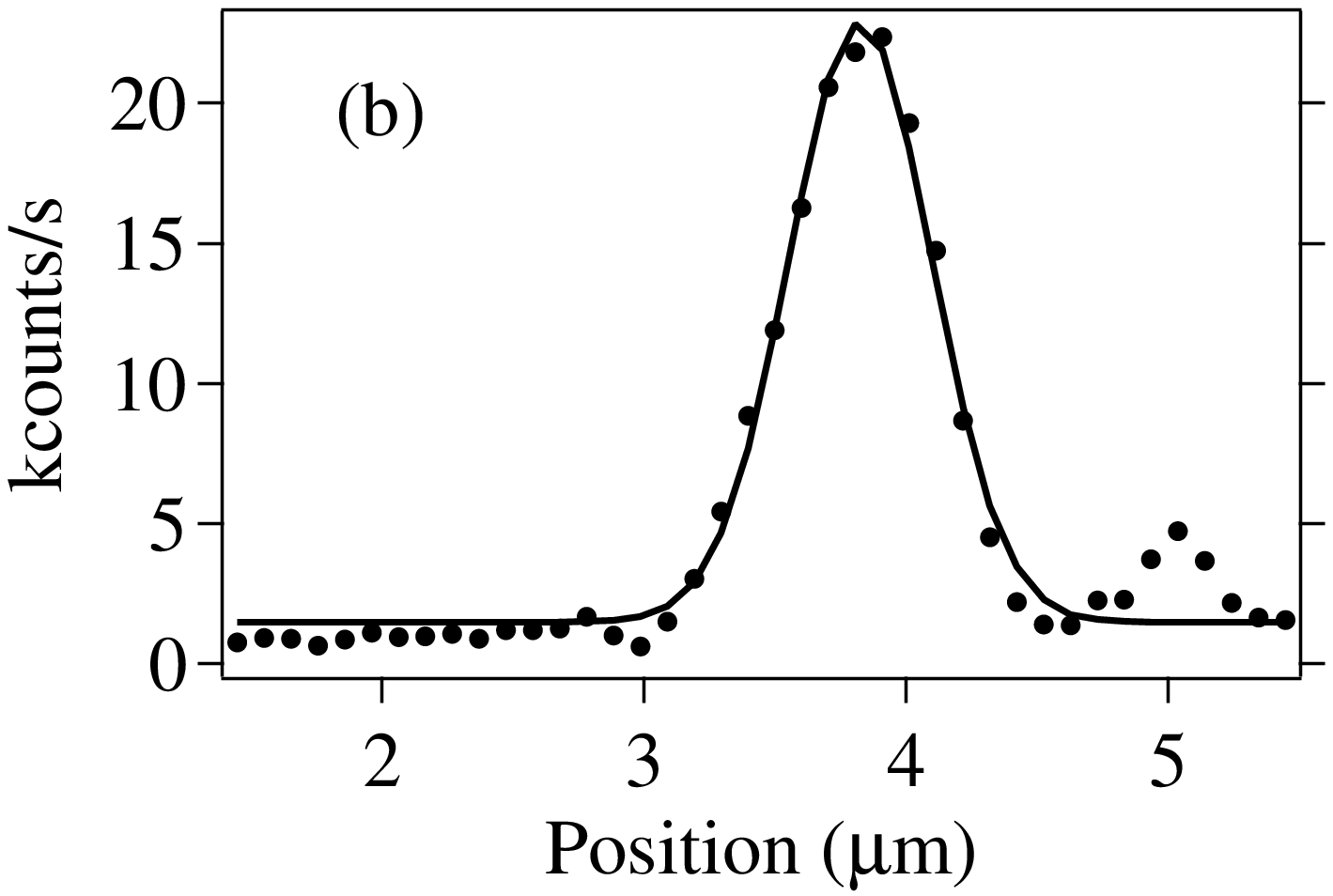}
\caption{(a) Confocal microscopy raster scan
($ 5 \; \times \;  5 \; \mu$m$^2$) of a diamond nanocrystal containing a
single NV center. The count rate corresponds to one detector only.
The size of a pixel is $100$ nm and the integration time per pixel is $32$ ms.
The laser intensity impinging on the sample is $2.7$ mW.
In (b) a linescan along the dotted line of (a) is displayed, together with
a gaussian fit, which is used to evaluate the signal and background levels.
Here we obtain $S/B=20$. Note that the fluorescence spot is slightly larger ($500$ nm)
than the
size of the excitation laser spot ($400$ nm) owing to saturation of the emitter.}
\label{scan}
\end{figure}

\begin{figure}
\includegraphics[scale=0.38]{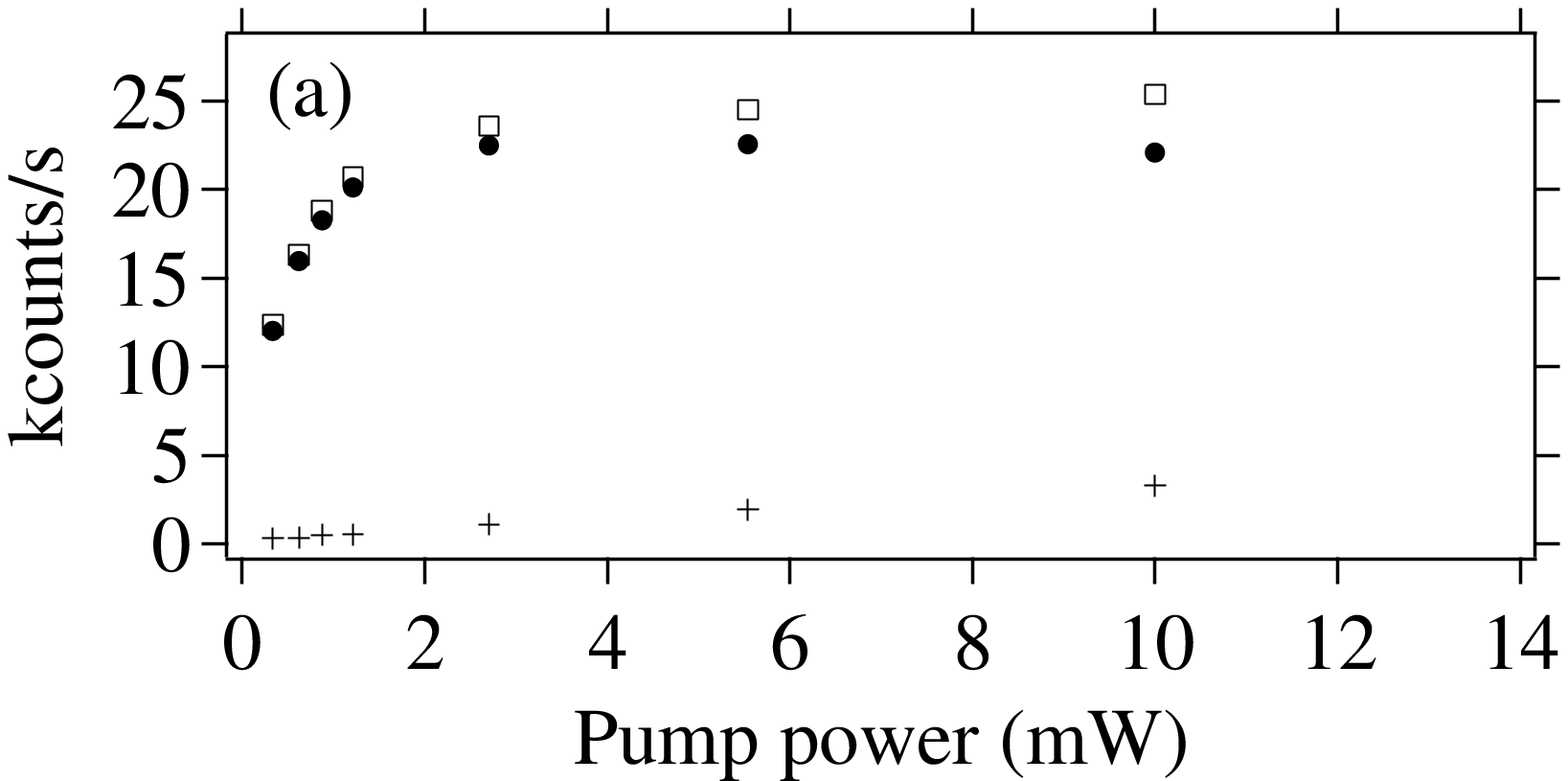}
\includegraphics[scale=0.38]{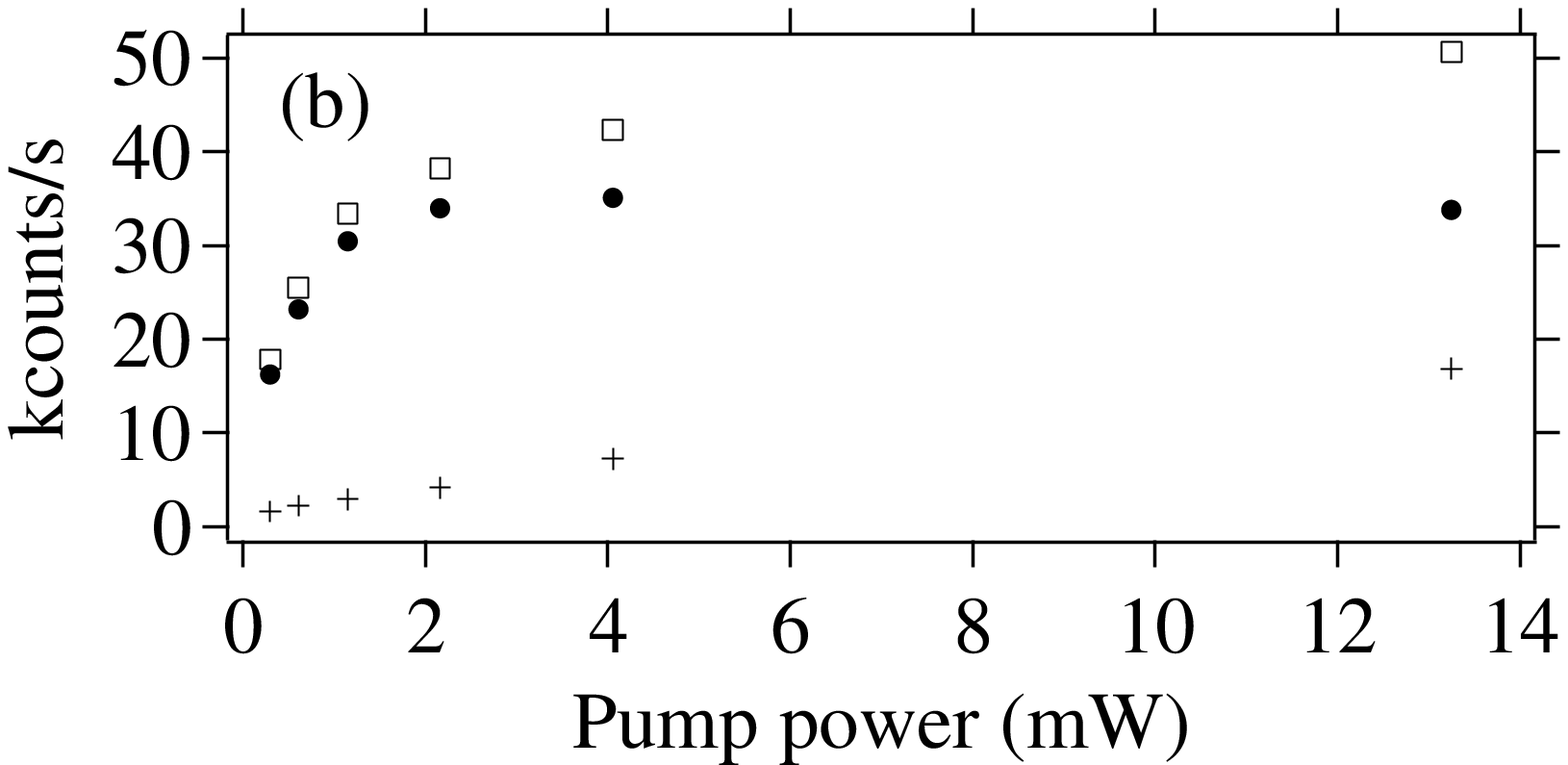}
\caption{Fluorescence rate  of a NV center in a nanocrystal (a) and in bulk
diamond (b)
as a function of the pump power.
The count rate corresponds to one detector only.
The crosses, empty squares, and black circles represent
the background $B$, the total count rate $T=S+B$,
and the signal from the center $S=T-B$, respectively.
The data for the nanocrystal corresponds to the same center as that shown in fig.
\protect\ref{scan}.}
\label{fsat}
\end{figure}

At first, we  investigate the fluorescence of single NV centers in
nanocrystals under CW
excitation with a frequency doubled YAG laser ($\lambda = 532$ nm). Fig. \ref{scan}(a)
displays a 2D scan of a single NV center.
From the line scan in fig.\ref{scan}(b) we deduce a spatial
resolution of $500$ nm and a signal (S) to
background (B) ratio of $S/B=20$. The corresponding value in the bulk 
is  $S/B=8$ (see fig. \ref{fsat}).

Fig. \ref{fsat} shows the saturation behavior of the
fluorescence rate of
NV centers in bulk and in nanocrystals with respect to pump
power.
The slightly decreasing rate at high pump power is due to a trapping
metastable state
\cite{capri}. The saturating count rate in a diamond nanocrystal 
($N_s^{nc}=4.4 \times 10^4$ s$^{-1}$)
 is slightly lower than that in bulk
diamond ($N_s^{b}=6.4 \times 10^4$  s$^{-1}$), but one has to take into account the
longer lifetime of a NV center in a nanocrystal (see at the end of this section). 
The  number of photons detected in a lifetime is
$\tau_{nc} N_s^{nc} = 11 \times 10^{-4}$ in the nanocrystal (lifetime $\tau_{nc}= 25$ ns)
and $\tau_{b} N_s^b = 7.4 \times 10^{-4}$ in  bulk diamond (lifetime $\tau_{b}= 11.6$ ns).
This means that the geometrical collection efficiency for nanocrystals 
is increased by $50\%$.

It can also be seen in fig. \ref{fsat} that the
contribution of background next to the nanocrystal is greatly reduced. 
The background coming from the diamond nanocrystal itself is reduced mainly
because the excited volume of diamond is smaller. 

\begin{figure}
\includegraphics[scale=0.38]{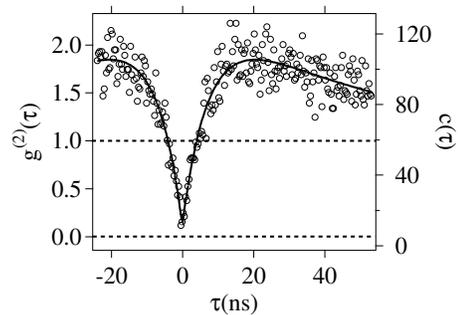}
\caption{Autocorrelation function $g^{(2)}(\tau )$ (left) and raw coincidence
rate (right) for a NV center in a nanocrystal.
The time bin $w=0.3$ ns,
total integration time is $323$ s and 
the laser intensity impinging on the sample is $2.7$ mW. Count rates on each 
photodiode are $N_1=22500$ s$^{-1}$ and $N_2=24500$ s$^{-1}$. The actual
number of  coincidences is indicated on the right. The zero-time value of the uncorrected
normalized correlation function given by eq. (\ref{CNCW}) is $C_N(0) = 0.17$.
The fit is performed with the model used in \protect\cite{capri}.
The data corresponds to the same center as that shown in fig. \protect\ref{scan}. }
\label{ABnc}
\end{figure}

The evaluated overall detection efficiency 
is given by $\eta_T^i= \eta_{geo}^i \eta_{opt}  \eta_{det}$, where $i=b,nc$ for the bulk or
for a nanocrystal, respectively. The geometrical collection efficiencies for a dipole oriented
orthogonally to the optical axis are $\eta_{geo}^b=0.18$ and $\eta_{geo}^{nc}=0.38$
(calculated). The optical transmission from the sample to the detectors is $\eta_{opt}=0.25$
(measured), and the detector quantum efficiency is $\eta_{det}=0.7$ (taken from the data
sheet). In addition the NV center presents a metastable state in which the excitation can be
shelved
\cite{GDTFWB,capri,KMZW}.
This leads to a reduction of the count rate by a factor equal to the saturated population
$\sigma _2^\infty=0.25 \pm 0.05$ of the excited state. This value
has been inferred by fitting a saturation curve (cf fig.
\ref{fsat}) and a set of autocorrelation functions (like fig. \ref{ABnc}) obtained for
different pump powers \cite{capri}. This fit involves many parameters and 
gives only approximative results.
 For a nanocrystal, the saturated count rate should then
be $S^{nc}_{cw}= \eta_T^{nc} \sigma _2^\infty /\tau_{nc} = 6.6\times10^5$ s$^{-1}$, which is 
$15$ times more than what we actually detected ($N_s=4.4 \times 10^4$ s$^{-1}$). This
discrepancy exists also in bulk and its origin is still under investigation \cite{efficiency}.

We have also studied the autocorrelation function of the fluorescence of single
NV centers in diamond
nanocrystals. The raw coincidences $c(\tau )$ (right axis) and autocorrelation
function $g^{(2)} (\tau )
= \langle I(t)I(t+ \tau ) \rangle / \langle I(t) \rangle ^2$ (left axis)
are represented in fig.
\ref{ABnc}.

For evaluating  the  intensity correlation function $g^{(2)}(\tau )$
of the NV center, the raw
correlation data $c(\tau )$ is normalized and corrected  in the following way.
The raw coincidence rate
$c(\tau)$ counted during a time $T$ within a time bin of width $w$ is first
normalized to that of a
Poissonian source according to the formula 
\beq
C_N^{cw}(\tau ) = c(\tau) /(N_1 N_2 w T) \; ,
\label{CNCW}
\eeq
 where $N_{1,2}$
are the count rates on each detector. The normalized coincidence rate
$C_N^{cw}(\tau ) $ is then
corrected for the background light $B$, and we obtain 
\beq
 g^{(2)}(\tau ) =
[C_N^{cw}(\tau ) - (1 - \rho^2)]/\rho^2 \; ,
\eeq
 where $\rho = S/(S+B)$ is related to the signal to background
ratio, which is measured independently in each experimental run by measuring the count rate
next to the nanocrystal (see fig. \ref{scan}(b)).  This takes into account only
the background coming from the fused silica and the polymer, and not the parasitic light
emitted by the diamond nanocrystal itself.  Note that we have checked experimentally that the
background light has a Poissonian statistics. 

It can be seen in fig. \ref{ABnc} that
$g^{(2)}(0) = 0.13$, where
the slight difference with zero is attributed to remaining background light
emitted by the nanocrystal and electronic jitter of the avalanche photodiodes ($300$ ps).
This almost vanishing value of $g^{(2)}(0 )$ is the signature of the presence of  a
single emitter in the observed nanocrystal. In the case of the presence of
$p$ centers within a nanocrystal, the value of the
zero-time antibunching is $1-1/p$. This is actually how we estimate the
number of NV centers in a
nanocrystal. 

We have obtained a $\tau = 0$ normalized coincidence rate $C_N^{cw}(0)=0.17$
 at the fluorescence rate maximum (input power of $2.7$
mW), where the best value
in bulk diamond was $0.26$  \cite{capri,KMZW}.
As we shall see in the next section, this uncorrected
normalized coincidence rate $C_N^{cw}(0)$ is the relevant parameter for
characterizing a single photon source.

It should  also be mentioned that $g^{(2)}(\tau )$ reaches
values greater than unity for $\be{\tau } \geq 10$ ns.
This bunching  effect for longer time scale  is due to the presence of the
trapping metastable state  in which the system can be shelved
\cite{KJRT,DFTJKNW,capri,BFTO}. This effect can also be seen as  blinking 
on a time scale of $\approx 50$ ns. The time distribution of photons can
ve viewed as bursts of photons of about $50$ ns duration. In each burst the delay between
successive photons is always larger than $5$ ns (antibunching).  

\begin{figure}
\includegraphics[scale=0.4]{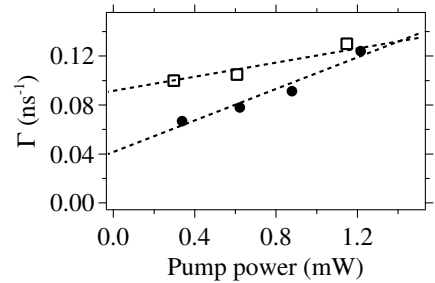}
\caption{Argument $\Gamma$ of the exponential function fitting the antibunching traces
obtained at different pump power. The extrapolation for vanishing pump power gives the
lifetime. The black circles correspond to the data of fig. \protect\ref{scan}
and yield $\tau _{nc}= 25$ ns.
The empty squares correspond to a NV center in bulk diamond ($\tau_b=11.6$ ns).
The slope for the nanocrystal is twice as large as that in bulk which
 is consistent with the lifetime increase, since the NV center in a nanocrystal
can absorb twice as many
photons during its lifetime.}
\label{ABtau}
\end{figure}

In a low pump regime,
the central dip in the antibunching traces can be fitted by an
exponential function with the argument $-\Gamma\be{\tau}$, where
$\Gamma= \gamma + r$, with $\gamma$ being
the NV center spontaneous decay rate
and $r$ the pumping rate \cite{capri}.
Such fits have been performed for different pumping powers.
The inverse lifetime $\gamma=1/\tau _{nc}$ of an NV center in a nanocrystal
can  then be inferred by  extrapolating the value of the time constant for
vanishing pump power (fig. \ref{ABtau}).
We deduce a lifetime for NV centers in bulk diamond of
$\tau_b = 11.6 \pm 0.1$ ns \cite{CTJ}
whereas  the lifetime is found to be $\tau_{nc} = 25 \pm 4$ ns
in diamond nanocrystals \cite{qph}.
This value has been obtained by
observing $10$ different nanocrystals. A possible explanation for this
lifetime increase is that the refractive
index experienced by the emitted light is different in bulk diamond and
nanocrystals.
When working out the spontaneous emission rate from the Fermi's golden rule,
it turns out that this rate is proportional to
 the refractive index $n$ of the material in which the
dipole is radiating.

In our case, the NV center in bulk diamond  emits within a medium of index
$n_d=2.4$, whereas the center
in a sub-wavelength nanocrystals emits as if it were in air for one half of
the space, and in fused silica
(refractive index $n_s=1.45$) for the other half.
The expected lifetime is then $\tau_{nc} = \tau_b [2 n_d/(1+n_s)] = 22.7$ ns
in good agreement with the experimental values.
A full description of lifetime changes due to refractive index is 
a controversial subject
 mainly because of local field correction issues
\cite{gl,cc,cb}.
However, our results tend to show that the local field experienced by
the NV center in bulk and in nanocrystals is
 the same \cite{qph}. As mentioned earlier no substantial difference in the
emission spectrum of NV centers
in nanocrystals and in bulk has been found. This is a good indication that 
the observed lifetime change is mainly due to the modification of the refractive index
of the medium in which the NV center is radiating.

\section{Pulsed excitation}

The pulsed excitation consists of a home build source along the lines 
of reference \cite{pct}.
The output of a $100$ mW  continuous
wave   Nd:YAG laser ($\lambda =1064$ nm) is coupled into a fast ($3$ GHz)
integrated LiNbO$_3$
modulator (Alenia) which slices up pulses of $1$ ns duration at a repetition
rate of $10$ MHz. The
pulses are then amplified to $1$ W mean power by an Ytterbium fiber
amplifier (Keopsys) and frequency
doubled using a PPKTP crystal. In this way we obtain  pulses at
$\lambda = 532$ ns of energy
$2.5$ nJ.

\begin{figure}
\includegraphics[scale=0.39]{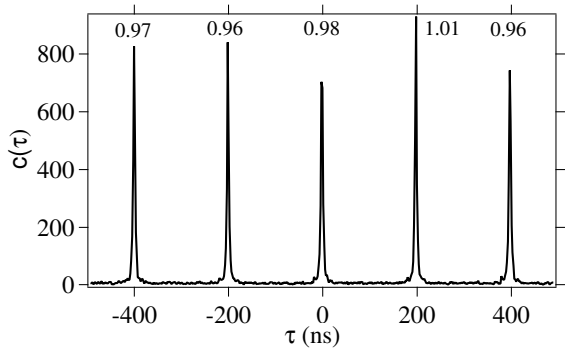}
\caption {Autocorrelation function from a fluorescing material (piece of white
paper) excited by  the pulsed laser. The emitted light has a Poissonian distribution. 
Pulse repetition period is $200$ ns and the pulse width 1.2 ns. The count rates are
$N_1 = 5011$ and $N_2 = 5343$ s$^{-1}$. Integration time is $T=595$
s and the time bin is
$2$ ns. The area of each individual peak normalizes to unity.}
\label{pg2}
\end{figure}

We  investigate the intensity autocorrelation function $g^{(2)}(\tau)$ of the
fluorescence light of a single NV center in a nanocrystal under pulsed
excitation.
 For a
Poissonian source the probability of having a coincidence between two
photons in the same pulse, or
two photons coming from different pulses, is equal.
Therefore the autocorrelation function for a
pulsed Poissonian source exhibits peaks of same height separated by the repetition period
(fig. \ref{pg2}).

\begin{figure}
\includegraphics[scale=0.39]{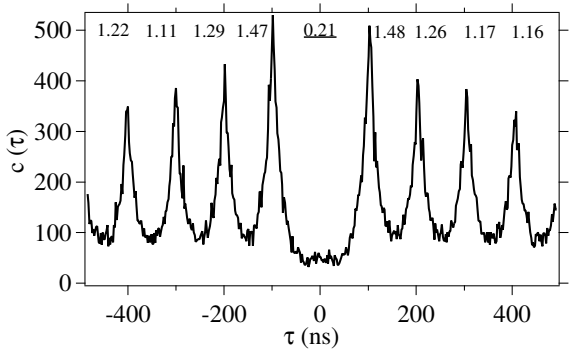}
\caption {Pulsed autocorrelation function of a single NV center. Pulse
repetition period is $100$ ns, pulse width $1.2$ ns and
excitation mean power $0.9$ mW. Count rates are $N_1 = 10504$ s$^{-1}$ and
$N_2=9995$ s$^{-1}$.
 Integration time is $T=588$ s and the time bin is $2$ ns. The
coincidences between peaks do not go down to zero because of the
overlapping of adjacent peaks. The number above each
peak represents its normalized area.}
\label{nvg2}
\end{figure}

In fig. \ref{nvg2} is shown the intensity autocorrelation function of a
single NV center under pulsed excitation. The
excitation pulse duration is $d=1.2$ ns and the repetition period is
$\theta=100$ ns. It can be seen that the peak around
$\tau =0$ is missing, which implies that the probability of having two photons
in one pulse is strongly reduced. This
gives rise to a highly sub-Poissonian light source.
In order to compare our single photon source to a pulsed Poissonian light
source, one has to normalize the area  $c(m)$ of peak number $m$ to the
area $N_1 N_2 \theta T$ of a peak corresponding to a pulsed Poissonian distribution with the
same count rate, with $T$ being the total acquisition time. Analogously to eq.(\ref{CNCW}),
the normalized area of each peak is given by
\beq
C_N(m)= c(m) / N_1 N_2 \theta T  \; .
\label{norm}
\eeq
The determination of the area $c(m)$ is performed in the following way.
The peaks are
 fitted by  exponential decays with the same lifetime. We checked that the lifetime
found from the fit is the same than that deduced from the CW excitation. 
This fitting procedure allows an accurate evaluation of the area $c(m)$ of  peak $m$,
in spite of the significant overlap between peaks.
 The normalized peak areas $C_N(m)$ are given by the numbers
displayed above each peak in fig. \ref{nvg2}. For a Poissonian pulsed light source
$C_N(m)=1$ for all $m$ (see fig. \ref{pg2}). For our single dipole in the most favorable case,
in which the NV center is saturated but the zero time peak is as low as possible, we obtained
$C_N(0)=0.21$. Note that this value is slightly larger than $C_N^{cw}(0)=0.17$.
This is attributed to the finite duration ($d=1.2$ ns) of the exciting pulses \cite{pra}.

Let us recall that the probability $p_2$  of having two photons in
a pulse is given by (assuming $p_2 \ll 1$)
\beq
p_2= C_N(0) \; p_1^2/2
\label{p2}
\eeq
where $p_1$  is the probability of having a single photon. Note that for
Poissonian light $C_N(0)=1$,
and eq.(\ref{p2}) with $C_N(0)=1$ gives the photon probability distribution of an
attenuated coherent pulse.
The zero time normalized coincidence rate $C_N(0)=0.21$
 means that the rate of two photon
pulses is nearly five times lower than for Poissonian light. 
Since our source has a single photon rate of $2\times 10^4$ s$^{-1}$
at an excitation repetition rate of $10$ MHz, its rate of two-photon pulses
is only of $4$ s$^{-1}$.

\begin{figure} 
\includegraphics[scale=0.4]{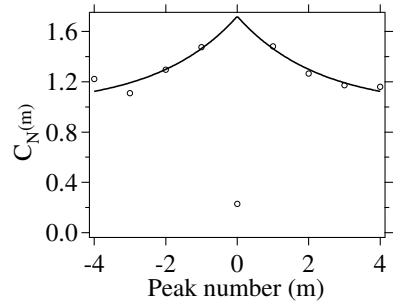}
\caption {Normalized peak area of fig. \protect\ref{nvg2} as function of
the peak number $m$. The experimental data is fitted with a function that
assumes a random blinking
of the NV center due to the trapping state.}
\label{tontoff}
\end{figure}

It can also be observed that the  peaks $m\neq 0$ grow above unity. 
 Just as in fig. \ref{ABnc}, this bunching effect for non zero time scale
comes from the existence of a
metastable state \cite{KJRT,GDTFWB,capri,KMZW,BFTO}.
In fig. \ref{tontoff} we have plotted the normalized area of the first peaks.
The long time decay behavior can be described by a simple model  assuming that the
NV center gets randomly trapped in the metastable state. This results in random blinking of
the NV center. The normalized area of peak $m$ is then given by \cite{YY}
\beq
C_N(m\neq 0)=1+\frac {T_{off}}{T_{on}} e^{-(1/T_{off}+1/T_{on}) |m| \theta} \; ,
\label {tonofffit}
\eeq
where $ T_{off}$ is
the mean time during which the excitation is trapped in the metastable state and
 the emission is inhibited and  $T_{on}$ is
the mean time during which the center is
emitting.
 Fitting the normalized peak area
(cf fig. \ref{tontoff}) with eq.(\ref{tonofffit}) allows us to extract the values
$T_{on}=460$ ns,  $T_{off}= 390$ ns. 
The saturated count rate should then be 
$S^{nc}_p = \eta^{nc}_T [T_{on}/(T_{on} + T_{off})]/ \theta = 1.4 \times 10^5$ s$^{-1}$, 
while we detect only $N_s=2\times 10^4$ s$^{-1}$. This is the same discrepancy that what
was observed for CW excitation \cite{efficiency}.
Note that the factor  $T_{on}/(T_{on} + T_{off})=0.54$ accounts for the shelving state and
plays the same role as $\sigma_2^{\infty}=0.25$ for CW excitation. Their different values
are attributed to the fact that the shelving and deshelving rates depend on the excitation
power
\cite{capri} and are therefore different in CW and in pulsed regime.

\section{Conclusion}

In this paper we have demonstrated the possibility of using single NV centers in
diamond nanocrystals as a room temperature solid state source for single photons. They
present the advantages of being photostable and easy to manipulate. Furthermore the
fabrication of the samples is easy and inexpensive. 
The single photon  rate summed over both photodiodes is
$2\times 10^4$ s$^{-1}$ for an excitation repetition rate of $10$ MHz. Our source exhibits a
strong sub-Poissonian distribution. The two photon pulse rate is reduced by a factor of five
compared to a Poissonian source and is equal to $4$  s$^{-1}$. Improvements in the collection
efficiency should be obtained by depositing the nanocrystals on a mirror or inserting them
into a microcavity. Even though the repetition rate is low compared to what can be obtained
with attenuated pulses using laser diodes, the reduction by a factor of 5 of the two photon
pulses will allow a secure transmission over larger distances \cite{L}.

\section{Acknowledgements}

We thank F. Treussard and R. Pansu for lending us crucial electronic
equipment, P. Georges for
constructive discussions on the pulsed laser system, E. Br\'eelle  from
the ``Groupe de
Physique des Solides'' at Paris 6 for the sample irradiation, and A. Machu
for sample annealing.
This work is supported by the European IST/FET program
``Quantum Information Processing and Telecommunication'',
project number 1999-10243 ``S4P". SK is supported by a Marie Curie fellowship from the
European Union.





\begin{references}
\bibitem{LoiPoisson} This approximation is valid if the mean photon number per pulse
is smaller than unity.
\end{references}
\end{document}